\documentclass{elsart5p}
\journal{Optics Communications}
\usepackage{amsfonts}
\usepackage{amssymb}
\usepackage{amsmath}
\usepackage{epsfig}
\usepackage{graphicx}
\usepackage{color}

\def\rootfig{./fig-ps/}

\begin{document}
%\linenumbers

\begin{frontmatter}
\title{Azimuthal Modulational Instability of Vortices in the
Nonlinear Schr{\"o}dinger Equation}

\author[sdsu]{R.M. Caplan},
\author[wnec]{Q.E. Hoq},
\author[sdsu]{R.\ Carretero-Gonz\'alez\corauthref{cor1}}, and
\corauth[cor1]{Corresponding author}
\ead[url]{http://www-rohan.sdsu.edu/$\sim$rcarrete/}
\author[pgk]{P.G.\ Kevrekidis}

\address[sdsu]{%
Nonlinear Dynamical Systems Group{$^1$},
Computational Sciences Research Center, and\\
Department of Mathematics and Statistics,
San Diego State University, San Diego,
CA 92182-7720, USA
}
\thanks[nlds]{{\tt URL:} http://nlds.sdsu.edu/}
\address[wnec]{%
Department of Mathematics, Western New England College,
Springfield, MA 01119
}
\address[pgk]{%
Department of Mathematics and Statistics, University of
Massachusetts, Amherst MA 01003-4515, USA}

\date{To appear in Optics Communications.}

%\documentclass[aps,floatfix,showpacs,preprintnumbers,twocolumn]{revtex4}
%\usepackage{amssymb}
%\usepackage{amsfonts}
%\usepackage{amsmath}
%\usepackage{graphicx}
%\usepackage{dcolumn}
%\usepackage{bm}
%
%\def\sn{{\rm sn}}
%\def\cn{{\rm cn}}
%\def\csch{{\rm csch}}
%
%\def\rootfig{./fig-ps/}
%
%\begin{document}
%
%\title{Azimuthal Modulational Instability of Vortices in the
%Nonlinear Schr{\"o}dinger Equation}
%%
%\author{R.M. Caplan$^{1}$}
%\author{Q.E. Hoq$^{2}$}
%\author{R. Carretero-Gonz\'{a}lez$^{1,}$\footnote{{\tt URL:} http://www.rohan.sdsu.edu/$\sim$rcarrete/}}
%\author{P.G. Kevrekidis$^{3}$}
%\affiliation{
%$^{1}$Nonlinear Dynamical Systems Group\footnote{{\tt URL:} http://nlds.sdsu.edu/}
%Computational Science Research Center\footnote{{\tt URL:} http://www.csrc.sdsu.edu/}, and
%Department of Mathematics and Statistics,
%San Diego State University, San Diego, California 92182-7720 USA \\
%$^{2}$ Department of Mathematics, Western New England College,
%Springfield, MA 01119    \\
%$^{3}$ Department of Mathematics and Statistics, University of Massachusetts,
%Amherst MA 01003-4515 \\
%}
%\date{\today}

\begin{abstract}
We study the azimuthal modulational instability of 
vortices with different topological charges, in the 
focusing two-dimensional nonlinear Schr{\"o}dinger 
(NLS) equation. The method of studying the stability 
relies on freezing the radial direction in the 
Lagrangian functional of the NLS in order to form a 
quasi-one-dimensional azimuthal equation of motion, 
and then applying a stability analysis in Fourier 
space of the azimuthal modes. We formulate predictions 
of growth rates of individual modes and find that 
vortices are unstable below a critical azimuthal wave 
number. Steady state vortex solutions are found by 
first using a variational approach to obtain an 
asymptotic analytical ansatz, and then using it as an 
initial condition to a numerical optimization routine.
The stability analysis predictions are corroborated by 
direct numerical simulations of the NLS. We briefly 
show how to extend the method to encompass nonlocal 
nonlinearities that tend to stabilize solutions.
\end{abstract}

\maketitle

%%use optional labels to link authors explicitly to addresses:
%%\author[label1,label2]{}
%%\address[label1]{}
%%\address[label2]{}

\begin{keyword}
Nonlinear optics \sep Nonlinear Schr\"{o}dinger equation \sep modulational
instability

\PACS
42.65.-k \sep% Nonlinear optics,
42.65.Sf \sep% Dynamics of nonlinear optical systems; optical instabilities, optical chaos and complexity, and optical spatio¿temporal dynamics
42.65.Jx \sep% Self-focusing in nonlinear optics,
03.75.Lm% Tunneling, Josephson effect, BECs in periodic potentials, solitons, vortices, and topological excitations
%42.70.Mp  % nonlinear optical crystals
\end{keyword}

\end{frontmatter}

%\pacs{
%42.65.-k, % Nonlinear optics,
%42.65.Sf, % Dynamics of nonlinear optical systems; optical instabilities, optical chaos and complexity, and optical spatio¿temporal dynamics
%42.65.Jx, % Self-focusing in nonlinear optics,
%03.75.Lm % Tunneling, Josephson effect, BECs in periodic potentials, solitons, vortices, and topological excitations
%%42.70.Mp  % nonlinear optical crystals
%}
%\maketitle

\section{Introduction}

%%%%%%%%%%%%%%

The  nonlinear Schr{\"o}dinger (NLS) equation has been used to describe a very
large variety of physical systems since it is the lowest order
{\em nonlinear} (cubic) partial differential equation that describes
the propagation of modulated waves \cite{sc03}.
%
%There are two main classes of NLS,
%depending on the sign of the nonlinearity.  These are the focusing
%(or attracting), and the defocusing (or repulsive) cases.  Which case
%to use depends on parameters in the physical system being described.
%
Two interesting systems described by the NLS that our study is relevant to are
Bose-Einstein Condensates (BECs) \cite{pethick,BECBOOK}
and
light propagation in amorphous optical media \cite{yura_agr}.

A BEC is an ultra-cold (on the order of $10^{-8} K$) gas of
$10^{3}$--$10^{6}$ atoms which have predominantly condensed into the
same quantum state, and therefore behaves like one large macroscopic
atom.  Its dynamics can be described (through a mean-field approach)
by a variant of the NLS called the Gross-Pitaevskii (GP) equation
that includes an external potential trapping the condensed atoms \cite{BECBOOK}:
\begin{equation}
\label{gp}
i\hbar\Psi_t = - \frac{\hbar^2}{2m_a} \nabla^2\Psi 
+ \frac{4\pi\hbar^2a_0}{m_a}|\Psi |^2\Psi
+ V_{\mbox{\scriptsize ext}}(\mathbf{r})\Psi,
\end{equation}
where $\hbar$ is the reduced Planck constant, $m_a$ is the mass of
one of the atoms in the condensate,
$V_{\mbox{\scriptsize ext}}(\mathbf{r})$ is the external potential,
$\nabla^2$ is the three-dimensional Laplacian, and
$a_0$ is the $s$-wave scattering length ($a_0<0$ corresponding
to the attractive [focusing] case while $a_0>0$ to the repulsive
[defocusing] case).
The modulus squared of the wave
function, $|\Psi|^2$, represents the density of the atoms in the
condensate.
In BECs, a focusing nonlinearity has the physical meaning that
the particles in the condensate will feature attractive
interactions. This can cause the BEC to collapse into
itself, which in turn increases the kinetic energy of the particles,
 and leads to an `explosive' destruction of the BEC dubbed a
`Bosenova' \cite{Sacket99,Gerton01,Donley01,BECcollapse}. 
In the defocusing case, the
particles have repulsive interactions, in which case
the BEC tries to expand (this is prevented by the external trap,
when the latter is present).
Although BECs are three-dimensional objects, by increasing
the strength of the external trap in one transverse direction,
one can reshape the BEC into a quasi-two-dimensional disk (or even
a quasi-one-dimensional cigar-shaped condensate in the case of
two strong transverse directions) \cite{BECov}. Each of these
situations can be described using appropriate forms of the
two-dimensional (2D) and one-dimensional GP equations
\cite{npse,Delgado:08,sc03,pethick,BECBOOK}.

On the other hand, amorphous optical media
such as silica exhibiting a Kerr effect can be modeled
using the NLS, where the modulus squared of
the wave function represents the intensity of the light
being propagated through the media.
In such a case, a $(2+1)$-dimensional NLS is used, where
the two dimensions of the wave function represent a
spatial cross-section,
while the third dimension $z$ (which represented time in the case of BECs)
represents the direction of propagation:
\begin{equation}
\label{NLScrystal}
2i\beta_0\Psi_z + \nabla^2\Psi + \beta_0^2\left(\frac{n_2}{n_0}\right)|\Psi|^2\Psi,
\end{equation}
where $\nabla^2$ is the 2D Laplacian, and $\beta_0$ is the
propagation constant.  The parameters $n_0$ and $n_2$
form the index of refraction in the medium as
$n = n_0+n_2|\Psi|^2$, where $n_0$ is the index of
refraction in the absence of light, and
$n_2$ is the change in the index of refraction due to the
intensity of the incident light  \cite{NLScrys}.

% The defocusing case here corresponds to
% a negative change in the refractive index of the crystal
% ($n_2<0$) which causes the light to defocus and spread out
% as it propagates through the crystal.  The focusing case
% ($n_2>0$), on the other hand,
% corresponds to a positive change in the refractive
% index of the crystal, which acts to focus the light,
% increasing its intensity. This focusing increases until
% the crystal is saturated, an effect which is not accounted
% for in Eq.~(\ref{NLScrystal}), but can be modeled using
% an NLS, e.g., with a saturable nonlinearity \cite{NLSsaturate}
% or by including higher order (such as quintic) terms \cite{yura_agr}.
% Despite this, the stability study presented here is still directly
% relevant because the saturation effects do not become
% important until the growth in intensity is very strong,
% and our study is limited to small perturbations.

The 2D NLS equation supports vortices \cite{Kruglov92}.
Vortices are ring-shaped structures which have a rotational
periodic angular phase associated to them.
A key property of the vortex is its
topological charge, denoted as $m$, which indicates how
many periods there are in the angular phase
around the vortex core \cite{BECBOOK}.
For $|m|>0$, the wave function at the center of the vortex
becomes identically zero, causing the ring-like shape.
As we will describe below,
vortex solutions of the NLS in the focusing case are
modulationally unstable in the azimuthal direction.
%Thus, a vortex will exhibit exponential growth
%of azimuthal modes, where each mode (denoted by an integer
%value $K$) has its own growth rate. This eventually
%leads to the collapse of the vortex into $K$ filaments.
%???
Our purpose in the present manuscript is to formulate
and test a method for studying the azimuthal modulational
instability \cite{Bigelow:02} of vortex solutions to the NLS.
The goal is to predict the growth
rates of the unstable modes, and predict the critical
mode, below which all modes are unstable.
Wherever relevant, we will make comparisons of the
semi-analytical methods presented herein to recent developments
in the study of ring vortices of the NLS equation, such
as Refs.~\cite{carr06,Herring08}.

The manuscript is organized as follows.
In Sec.~\ref{model}, using the Lagrangian representation
of the NLS, we formulate a quasi-one-dimensional equation
of motion for the dynamics of separable steady-state
vortex solutions to the NLS.  Then, in Sec.~\ref{stable}
we describe the azimuthal modulational stability
analysis yielding predictions of the
growth rates of unstable modes, as well as the critical
mode, below which all modes are unstable.
In Sec.~\ref{sec_VA} a variational approach is used to
obtain a reliable ansatz for the radial profile of steady-state
vortices.
In Sec.~\ref{SEC:num} the
ansatz is refined into a numerically `exact' radial
profile using optimization methods and then numerically
integrated to extract azimuthal growth rates
and critical modes that are found
to match well to our analytical predictions. In Sec.~\ref{sec_nl} we
show an extension of the technique to the NLS with {\em nonlocal}
nonlinearity \cite{krol04}.
Finally, in Sec.~\ref{SEC:CONCLU} we summarize our results and
give some concluding remarks.

\section{Azimuthal Equation of Motion}
\label{model}

Both physical scenarios described above (BECs and amorphous optical media)
can be modeled, under appropriate conditions, by the
2D NLS.
Let us then use the non-dimensionalized NLS
\begin{equation}
i\Psi_t+\nabla^2\Psi+
s\left|\Psi\right|^2\!\Psi=0,
\label{nls}
\end{equation}
where $\nabla^2\Psi$ is the 2D Laplacian of the
wave function $\Psi$ and $s=+1$ ($s=-1$) denotes the focusing
(defocusing) case. The action functional of Eq.~(\ref{nls}) is:
\begin{equation}
S=\int_0^\infty \!\!L\,dt,
\label{acfunc}
\end{equation}
where the Lagrangian reads
\begin{equation}
L = \int_0^{2\pi}\!\!\int_0^\infty\!
    \mathcal{L}\,r\,dr\,d\theta,
\label{lag}
\end{equation}
and its Lagrangian density, in polar coordinates, corresponds to \cite{VA}
\begin{equation}
\mathcal{L}=\frac{i}{2}\left(\Psi\Psi_t^* -
             \Psi^*\Psi_t\right) +
             \left|\Psi_r + \frac{1}{r}\Psi_\theta\right|^2
             - \frac{s}{2}|\Psi|^4.
\label{lagden}
\end{equation}

In order to find the azimuthal equation of motion, we assume a separable solution with a steady-state ``frozen" in time radial profile:
\begin{equation}
\label{sepAf}
\Psi(r,\theta,t)=f(r)\,A(\theta,t),
\end{equation}
where all of the phase components of the solution are contained in $A$,
and therefore $f(r)\in\mathbb{R}$.  It is worth mentioning that vortex
solutions to Eq.~(\ref{nls}) are not necessarily completely separable
as per Eq.~(\ref{sepAf}) and thus this property needs to be checked
(see Sec.~\ref{sec_sim} for more details).

When Eq.~(\ref{sepAf}) is inserted into Eq.~(\ref{lag}), since
$f(r)$ is ``frozen'', all radial integrals of Eq.~(\ref{lag})
become constants. This allows us to transform the 2D Lagrangian
into a quasi-one dimensional (in $\theta$) Lagrangian which can be
used to find the equation of motion for $A(\theta,t)$.  We use the
term `quasi-one-dimensional' because although it becomes a
one-dimensional problem, the radial direction is not ignored, but
shows implicitly in the values of the radial integral constants.

First, we insert Eq.~(\ref{sepAf}) into the Lagrangian density and evaluate the radial integrals of the Lagrangian to obtain our quasi-one-dimensional Lagrangian density:
\begin{alignat}{2}
\label{l1d}
\mathcal{L}_{\mbox{\scriptsize 1D}} =&\frac{i}{2}C_1(AA_t^* - A^*A_t)
 + C_2|A|^2 + C_3|A_\theta|^2
\\
&+C_5A_\theta^*A +
C_6A_\theta A^* - \frac{s}{2}C_4|A|^4,
\notag
\end{alignat}
where
\begin{alignat}{3}
\label{constants}
C_1&= \int_0^\infty \left|f(r)\right|^2r\,dr,  &~~ C_2&= \int_0^\infty \left|\frac{df}{dr}\right|^2r\,dr,
\\[1.0ex]
C_3&=\int_0^\infty\!\frac{1}{r^2}\left|f(r)\right|^2r\,dr, &~~ C_4&=\int_0^\infty\!\left|f(r)\right|^4r\,dr,
\notag \\[1.0ex]
C_5&= \int_0^\infty\!\frac{1}{r}\frac{df}{dr}f(r)^*\,r\,dr, &~~ C_6&= \int_0^\infty\!\frac{1}{r}\!\left(\frac{df}{dr}\right)^{\!\!*}\!f(r)\,r\,dr.
\notag
\end{alignat}

We evaluate the variational derivative of the action functional as shown in Ref.~\cite{VA}, which in this case takes the form:
\begin{equation}
\label{varder}
\frac{\delta S}{\delta A^*}=
\frac{\partial}{\partial t}
\frac{\partial \mathcal{L}_{\mbox{\scriptsize 1D}}}{\partial \left[A_t^*\right]}+
\frac{\partial}{\partial \theta}
\frac{\partial \mathcal{L}_{\mbox{\scriptsize 1D}}}{\partial \left[A_\theta^*\right]}
-\frac{\partial \mathcal{L}_{\mbox{\scriptsize 1D}}}{\partial A^*} =0.
\end{equation}
Inserting Eq.~(\ref{l1d}) into Eq.~(\ref{varder}) yields the evolution equation for $A(\theta,t)$:
\[
i\,C_1A_t = C_2A -
C_3A_{\theta\theta} + (C_5 - C_6)A_\theta -
s\,C_4|A|^2A.
\]
Since $f(r)$ is real-valued, and $C_5 = C_6^*$, the $A_{\theta}$ term drops out:
\begin{equation}
i\,C_1A_t = C_2A - C_3A_{\theta\theta} - s\,C_4|A|^2A. \label{Aeq0}
\end{equation}
Applying the rescalings
\begin{equation}
\label{Arescale}
A\rightarrow A\,\mbox{exp}\left(-i\frac{C_2}{C_1} t\right),
\end{equation}
and
\begin{equation}
\label{trescale}
t\rightarrow \frac{C_3}{C_1}\,t,
\end{equation}
yields the azimuthal NLS
\begin{equation}
\label{Aeq}
iA_t=-A_{\theta\theta}-s\frac{C_4}{C_3}|A|^2A,
\end{equation}
that we next study for its modulational instability.

\section{Stability Analysis}
\label{stable}

For the stability analysis, we assume a
vortex solution of Eq.~(\ref{Aeq}):
\begin{equation}
\label{pwave}
A(\theta,t) = e^{i(m\theta +\Omega^{'} t)},
\end{equation}
where $m$ is the topological charge of the vortex, and $\Omega^{'}$ is the
frequency of rotation of the complex phase.  Notice that in this context,
the vortex waveform becomes an ``azimuthal plane wave'', and as such
its stability analysis becomes the standard modulational stability
analysis of this plane wave (which we briefly review for completeness
purposes here) \cite{hasegawa}.
The amplitude of the plane wave does not appear as an explicit term because it is absorbed into the radial profile $f(r)$ of Eq.~(\ref{sepAf}).  Inserting Eq.~(\ref{pwave}) into Eq.~(\ref{Aeq}), we get the following dispersion relation:
\begin{equation}
\label{disprel}
\Omega^{'} = -m^2 + s\frac{C_4}{C_3}.
\end{equation}

Let us now derive equations of motion for a complex perturbation.  Specifically, we wish to derive the amplitude equations for each perturbed Fourier mode.  We start by perturbing Eq.~(\ref{pwave}) with a complex, time-dependent perturbation of the form:
\begin{equation}
\label{Apert}
A(\theta,t) = \left(1 + u(\theta,t) + iv(\theta,t)\right)\,e^{i(m\theta + \Omega^{'} t)},
\end{equation}
where $|u|,|v| \ll 1$.

If we rescale time according to the rotating vorticity frame as:
\[
\tau = t + \frac{1}{2m}\,\theta,
\]
this yields
\begin{alignat}{2}
\label{uvPDE2}
u_t &= - v_{\theta\theta}
 - \left[s\frac{C_4}{C_3}\left(2uv +
u^2 v + v^3\right)\right],
\\[1.0ex]
v_t &=  u_{\theta\theta}
 + 2s\frac{C_4}{C_3}\,u +
\left[s\frac{C_4}{C_3}\left(v^2 + 3u^2
 + v^2u + u^3\right)\right]. \notag
\end{alignat}

As in Refs.~\cite{hasegawa,MI1D}, in order to study
modulational instability (MI), we seek amplitude
equations for the azimuthal modes by expanding $u$ and $v$ in a discrete
Fourier series:
\begin{alignat}{2}
\label{uvexpand}
u(\theta,t)&=\frac{1}{2\pi}\sum_{K=-\infty}^{\infty}\hat u (K,t)
 e^{-iK\theta}, \\[1.0ex]
v(\theta,t)&=\frac{1}{2\pi}\sum_{K=-\infty}^{\infty}\hat v (K,t)
 e^{-iK\theta},\notag
\end{alignat}
where $K$ is the mode number and its respective amplitude is given by:
\begin{alignat}{2}
\label{uvtrans}
\hat u(K,t) &= \int_0^{2\pi}\!u(\theta,t)\,e^{iK\theta}\,d\theta,
\\[1.0ex]
\hat v(K,t) &= \int_0^{2\pi}\!v(\theta,t)\,e^{iK\theta}\,d\theta.\notag
\end{alignat}
Applying these to Eq.~(\ref{uvPDE2}) yields two coupled nonlinear ordinary differential equations describing the dynamics for the amplitudes of $u$ and $v$ for each mode.  Since we are not interested in the long-term dynamics of the system, but only in the MI of small perturbations, we drop the nonlinear terms and write the resulting linearized system in matrix form as:
\begin{equation}
\label{muv}
\frac{d}{dt}\left[ \begin{array}{c}
\hat u \\[2.0ex] \hat v \end{array}\right]
 = \left[ \begin{array}{cc}
0 & K^2 \\[0.5ex]
\left(2s\frac{C_4}{C_3} - K^2\right) &
 0
\end{array}\right]\!
\left[ \begin{array}{c} \hat u \\[2.0ex] \hat v\end{array}
\right].
\end{equation}
The eigenvalues for this linear system are:
\[
\lambda_{\pm} = \pm \sqrt{K^2\left(2s\frac{C_4}{C_3}-K^2\right)}.
\]
We notice that for the defocusing nonlinearity ($s=-1$) {\em dark}
vortices are supported by a non-zero background, and thus
the $C_i$ integrals do not converge, and therefore the
method employed above would need to be adjusted by appropriately
subtracting the background field in the Lagrangian integrals.
%, since
%$C_4$, $C_3 \ge 0$, the eigenvalues are purely imaginary and
%therefore all small perturbations are neutrally stable and
%thus `dark' vortices are predicted to be azimuthally
%modulationally {\em stable}.
%
Nonetheless, it is worth mentioning that higher ($m>1$) charge
dark vortices are unstable since they tend to split into unitary
charge vortices as shown in Ref.~\cite{Ginzburg:58}
(see also Refs.~\cite{Pu99,Kawaguchi:04,Okano07,Isoshima07}, and references
therein, for recent work on this topic).
However, this instability is not
of the modulational type and thus we do not study it here,
and therefore we concentrate on the focusing case of $s=+1$ (`bright'
vortices) below.
It is also interesting to note that the presence
of a confining potential
might stabilize higher order dark vortices in certain
parameter windows \cite{Pu99,Mihalache:06,Alexander:02,Herring08}.

\begin{figure}[htb]
\centering
\includegraphics[width=4cm]{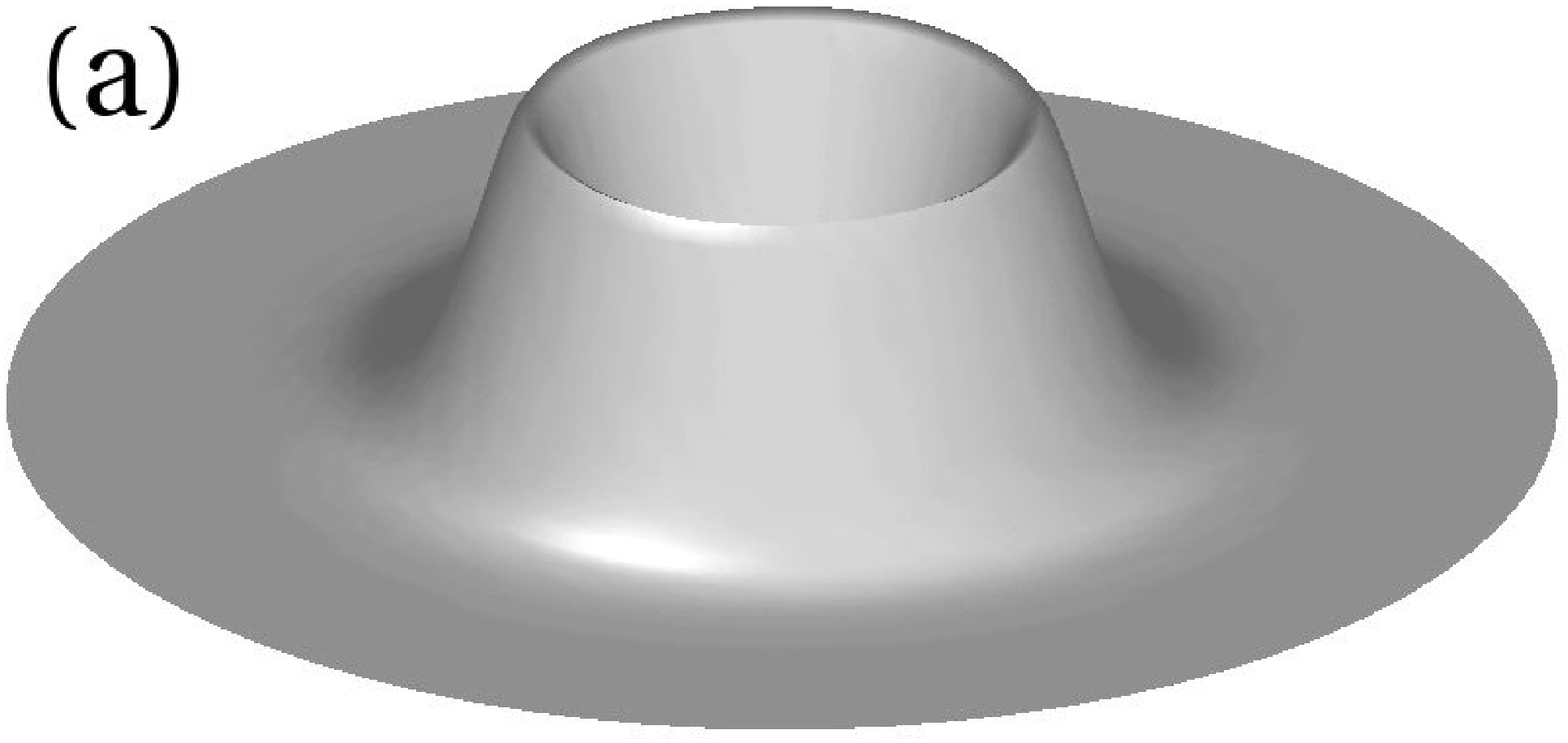} ~
\includegraphics[width=4cm]{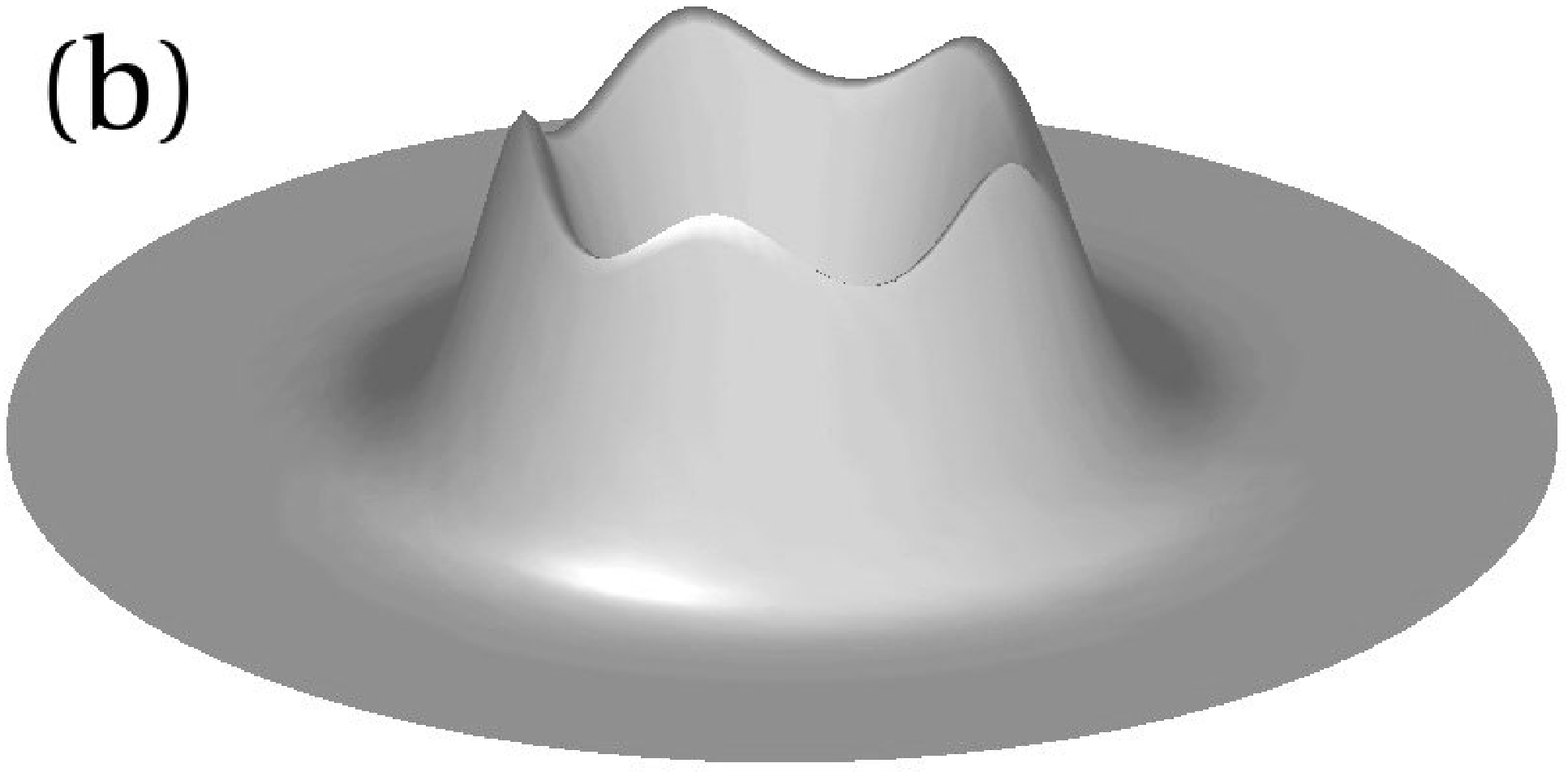} \\
\includegraphics[width=4cm]{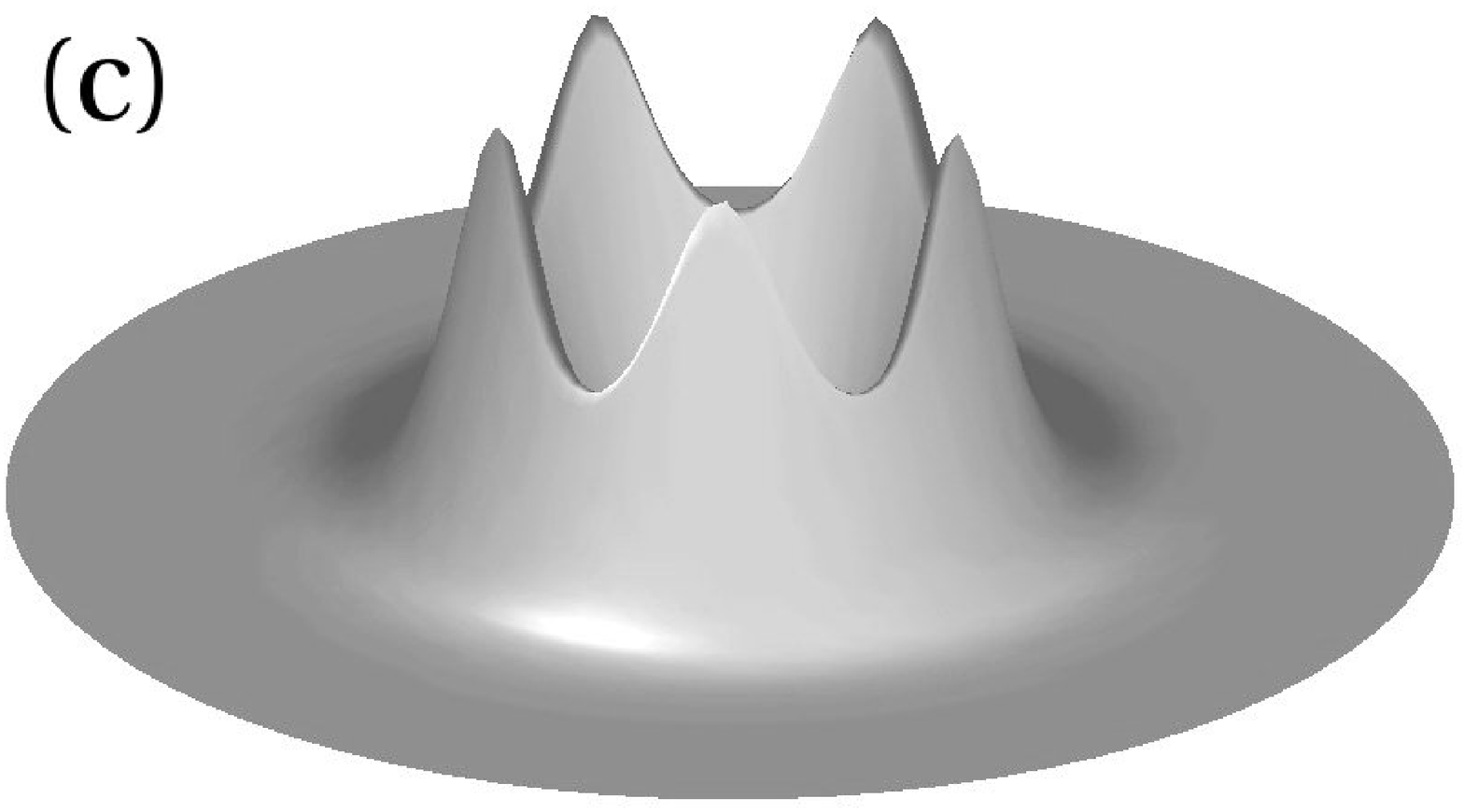} ~
\includegraphics[width=4cm]{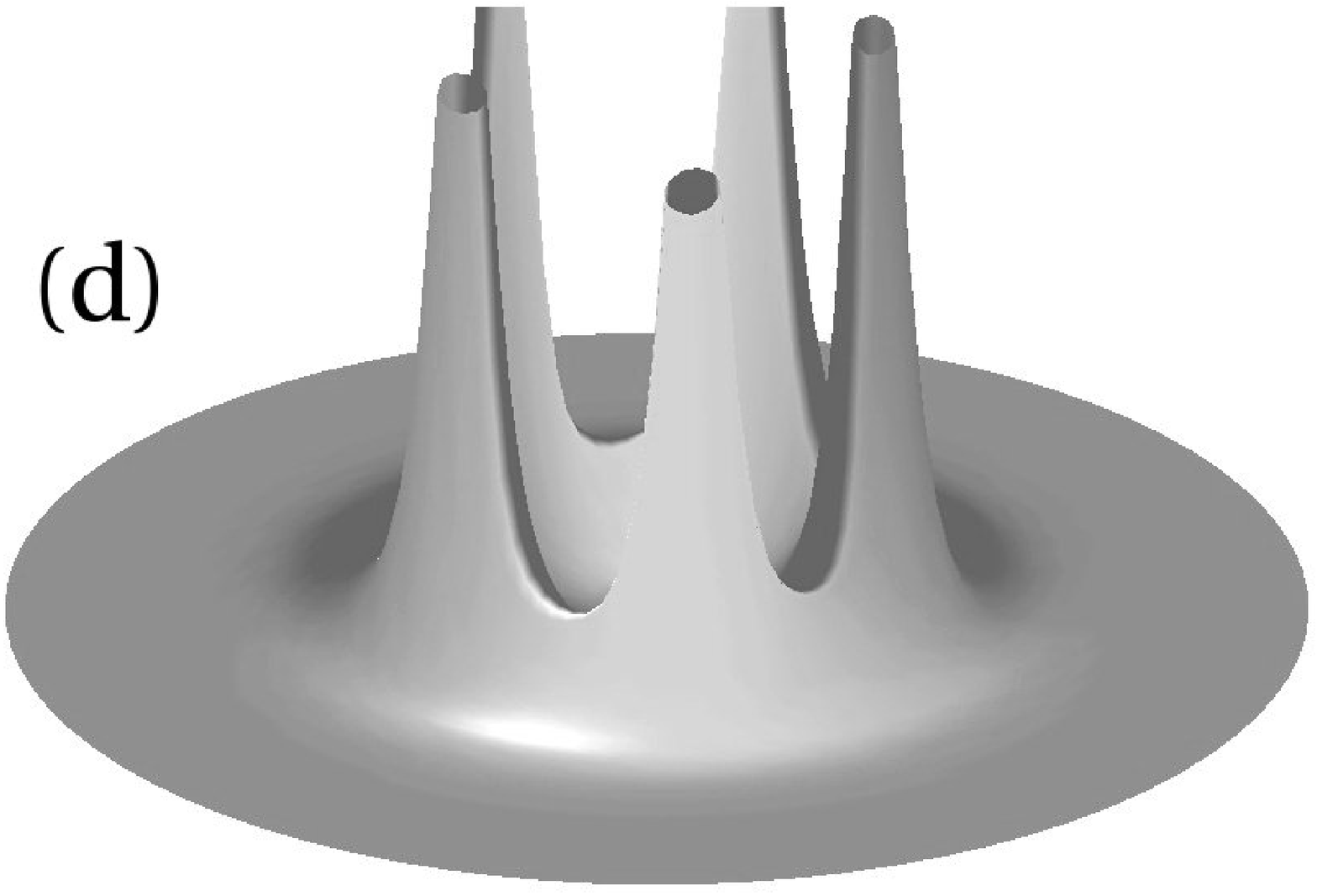}
%\vspace{-0.2cm}
\caption[Example of MI]{A typical numerical simulation of a
vortex solution to the NLS showing MI.
The vortex shown is of charge $m=2$ perturbed with mode $K=5$ starting
with a perturbation amplitude of $\epsilon=0.001$.
(a) $t=0$, (b) $t=8$, (c) $t=10$, and (d) $t=12$.
\label{MI_ex}}
\end{figure}

Returning to the case of interest, namely the focusing case ($s=+1$),
there is a bifurcation at a critical value of $K$:
\begin{equation}
\label{kcrit}
K_{\mbox{\scriptsize crit}} \equiv \sqrt{2s\frac{C_4}{C_3}},
\end{equation}
where $K<K_{\mbox{\scriptsize crit}}$ indicates a modulational instability.  An example of such MI is shown in Fig.~\ref{MI_ex}.

To predict the actual exponential 
growth rates for the perturbation of each mode from the eigenvalues, the time rescaling of Eq.~(\ref{trescale}) needs to be taken into account, in which case the growth rates (in terms of $K_{\mbox{\scriptsize crit}}$) are:
\begin{equation}
\label{eig2}
\lambda_{\pm} = \pm\frac{C_3}{C_1}\sqrt{K^2
\left(K_{\mbox{\scriptsize crit}}^2 - K^2 \right)}.
\end{equation}

\section{Variational Approach}
\label{sec_VA}

Explicit solutions for two dimensional steady-state vortices of the NLS are not available.  Therefore, in order to find a tractable, approximate, solution, we use a variational approach (VA) to get a reasonable ansatz, and then use that ansatz as an initial condition to a nonlinear equation optimization routine which finds the numerically `exact' steady-state profile, $f(r)$.
The VA-inferred seed may also be of value as an initial guess
to other numerical techniques that have
been previously used to obtain such vortices, including shooting
methods \cite{carr06} or Newton-type, fixed point schemes \cite{Herring08}.
The modified Gauss-Newton scheme presented below is intended as an
alternative to the former ones.

To perform the VA, we use the technique described in Ref.~\cite{VA}.  We insert a vortex ansatz with variable parameters into the Lagrangian of the NLS, and use the Euler-Lagrange equations to find the `best' values for the parameters.  We start with a general, separable, steady-state solution:
\begin{equation}
\label{var_sep}
\Psi(r,\theta,t) = f(r)e^{i(m\theta+\Omega t)},
\end{equation}
where $f(r)$ is the steady-state radial profile which we want to find.  Inserting this solution into the Lagrangian density of the NLS yields:
\begin{equation}
\label{var_langC}
L = 2\pi\left(\Omega \, C_1 + m^2\, C_3 + C_2 - \frac{1}{2}\, C_4\right),
\end{equation}
where we have now explicitly set $s=+1$ and the $C$-constants are the same as in Eq.~(\ref{constants}).

We use a one-dimensional soliton sech ansatz similar to that used in Ref.~\cite{SECH}:
\begin{equation}
\label{sech_ansatz}
f(r) = \sqrt{B}\,\mbox{sech}\left(\sqrt{\frac{B}{2}}(r-r_c)\right),
\end{equation}
with parameters $B$ and $r_c$ corresponding, respectively, to the
amplitude and location of the ring induced by the vortex.
Now assuming $r_c$ to be large, we can approximate the $C$-constants as follows:
\begin{eqnarray}
C_1 &=& \int_0^\infty\! B\,\mbox{sech}^2\left(F\right)r\,dr 
%\notag
%\\
%&=& 
=2\,\mbox{ln}\left(1+E \right)
\approx r_c\,\sqrt{8B}, 
%\approx 2\,\sqrt{2}r_c\,B^{1/2}, 
\label{VA2C1}
\\[2.0ex]
C_2 &=& \int_0^\infty\! \frac{B^2}{2}\,\mbox{sech}^2\left(F\right)\,\mbox{tanh}^2\left(F\right)r\,dr 
%\notag
\\
&=&\frac{B\left[\ln\left(1+E  \right)\left(E^2+E +1\right)+ 2\,E \right]}{3\left( 1+E \right) ^2}
%%\notag
%%\\
%%&\approx& 
\approx\frac{\sqrt{2}}{3}r_c\,B^{3/2}.
\notag
\\[2.0ex]
C_4 &=& \int_0^\infty\! B^2\,\mbox{sech}^4\left(F\right)r\,dr =4\,C_2 \label{VA2C4} 
%\notag
%\\
%&\approx& 
\approx\frac{4\sqrt{2}}{3}r_c\,B^{3/2},
\end{eqnarray}
where $E\equiv e^{\sqrt{2B}r_c}$ and $F\equiv \sqrt{\frac{B}{2}}(r-r_c)$.
The $C_3$ integral does not converge due to the singularity at $r=0$.
However, since we assume $r_c$ to be large, the $r$ in the integrand
can be viewed as a constant (which we choose to be the center of the
sech, i.e. $r_c$) and so we have:
\begin{eqnarray}
C_3 &=&  \int_0^\infty\! \frac{1}{r}\,B\,\mbox{sech}^2\left(F\right)\,dr
%\notag
%\\
%&\approx&
\approx\frac{1}{r_c} \int_0^\infty\! B\,\mbox{sech}^2\left(F\right)\,dr  \notag
\\
&=&  \frac{\sqrt{8B}}{r_c}\,\frac{e^{\sqrt{2B}r_c}}{(1+E )}
%\notag \\ &\approx&
\approx \frac{\sqrt{8B}}{r_c}, \label{VA2C3}
\end{eqnarray}
Using these approximations, we obtain:
\[
L = \frac{2\pi}{3}\,\sqrt{2B}\,r_c\left(6\,\Omega + 6\,\frac{m^2}{r_c^2}-B\right).
\]
The Euler-Lagrange equations:
\[
\frac{\partial L}{\partial B}=0, \qquad
\frac{\partial L}{\partial r_c}=0,
\]
lead us to the solution:
\begin{equation}
\label{VA2params}
B = 3 \Omega, \qquad r_c = \sqrt{\frac{2m^2}{\Omega}},
\end{equation}
and so our VA ansatz is:
\begin{equation}
\label{ne_sol}
f(r) = \sqrt{3\,\Omega}\,\mbox{sech}\left[\sqrt{\frac{3\,\Omega}{2}}\left(r-\sqrt{\frac{2m^2}{\Omega}}\right)\right].
\end{equation}

Despite the obvious problem with this solution at $r=0$ (where $f(0)$ should identically be equal to $0$), it captures the shape and position of the numerically `exact' solution very well (see Fig.~\ref{AnyvsNum}).  Also for higher $m$, its value at $r=0$ becomes very close to zero.

Using the VA ansatz with the asymptotic approximations of Eqs.~(\ref{VA2C1})--(\ref{VA2C3}), we can calculate analytical expressions for the exponential 
growth rates and critical modes of the MI:
\begin{alignat}{2}
K_{\mbox{\scriptsize crit}}^{\mbox{\scriptsize va}} &= \pm 2 \sqrt{2}m, \label{VA2pred}
\\[1.0ex]
\lambda^{\mbox{\scriptsize va}} &= K\Omega\, \frac{\sqrt{8m^2 - K^2}}{2m^2}. \notag
\end{alignat}
The advantage of the above formula is that, although approximate,
they describe in simple terms the MI experienced by the
vortex. Also, it should be noted that this analytical prediction
becomes more accurate for higher order vortices as the VA
is able to closely match the actual solution as depicted
in Fig.~\ref{AnyvsNum}.

\section{Numerical Results\label{SEC:num}}
\subsection{Numerical Optimization}
\label{sec_opti}

\begin{figure}[htb]
\centering
\includegraphics[width=8.5cm]{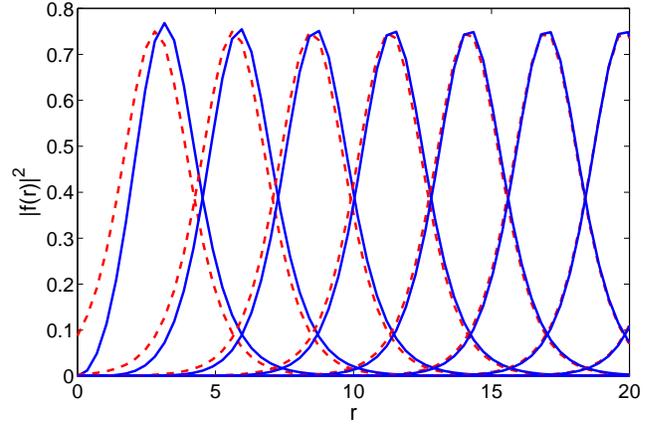}
\vspace{-0.2cm}
\caption[Comparison between VA and numerically exact solutions]{
(Color online) Comparison between the VA ansatz (dashed/red lines) and
the numerically `exact' solution (solid/blue lines) for charges
$m=1,...,7$ (curves left to right). We notice that the VA captures
the numerically `exact' solution very well as $m$ increases.
\label{AnyvsNum}}
\end{figure}

To refine the ansatz profile into a numerically `exact' solution,
we implement a nonlinear optimization scheme based on a modified Gauss-Newton
scheme \cite{Noce}.
First, we insert the following separable steady-state solution into Eq.~(\ref{nls}):
\begin{equation}
\label{fullsepsol}
\Psi(r,\theta,t)=f(r)\,e^{i(m\theta+\Omega t)},
\end{equation}
which produces an ordinary differential equation for $f(r)$ 
which can be discretized as:
\begin{equation}
\label{descODE}
F_i(f_i(r_i))=
-\left(\Omega+\frac{m^2}{r_i^2}\right) f_i + D(f_i) + f_i^3 = 0,
\end{equation}
where
\begin{equation}
\label{descODEOP}
D(f_i) =
\frac{1}{r_i}
\frac{1}{\Delta r}
\left(r_{i+\frac{1}{2}}\frac{f_{i+1}-f_{i}}{\Delta r}
 - r_{i-\frac{1}{2}}\frac{f_{i}-f_{i-1}}{\Delta r}\right).
\end{equation}
We now want a profile, $\vec f_0$, which optimizes $\vec F$ towards
the specific value $\vec 0$.
To do this we iterate the trial profile using:
\[
\vec f_{k+1} = \vec f_k + \alpha_k \vec p_k,
\]
where the step length $\alpha_k$, is found by:
\[
\min_{\alpha >0}M(\vec f_k + \alpha \vec p_k)
\rightarrow \alpha_k,
\]
and where $M(\vec f_k)$ is the merit function defined by:
\begin{equation}
\label{merit}
M(\vec f) = \frac{1}{2}\sum_{i=1}^n\,(F_i(\vec f))^2.
\end{equation}
The step direction, $\vec p_k$ is found using a modified Gauss-Newton (GN) formulation:
\begin{equation}
\label{GNstep}
\vec p_k = -(J_k^TJ_k+\Lambda_k I)^{-1}J_k^T\vec F(f_k),
\end{equation}
where $\Lambda_k$ is called the forcing term, which ensures that the step is always defined, even near non-zero roots of $M$.  The forcing term is manually set to values which produce desired results for our problem ($\Lambda_k=0.001$).  Some sample profiles for various charges are shown in Fig.~\ref{AnyvsNum}, where we can see the very good agreement between VA and the numerically `exact' solution, especially for higher charges.

The apparent convergence of the VA ansatz with the GN refined profile as
$|m|$
%>\infty$
increases can be very useful.  For very large $|m|$, the GN computation using a high enough resolution to avoid numerical errors can become very computationally expensive.  Therefore, the analytic stability predictions of the VA [see Eq.~(\ref{VA2pred})] can be used for predictions without the need to run numerical computations at all.  Even for low charges, the VA ansatz accurately describes the radius and maximum intensity of the vortex.

\begin{figure}[htb]
\centering
\includegraphics[width=8.5cm]{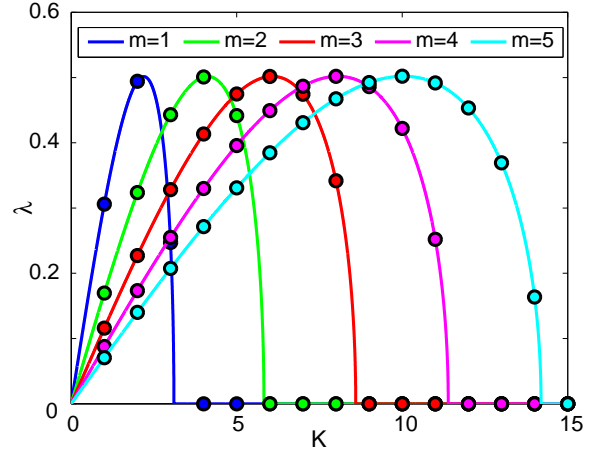}
\vspace{-0.2cm}
\caption[Numerical predictions of growth rates for $m=1,...,5$ using GN optimization.]
{(Color online) Numerical predictions of growth rates of perturbations 
of azimuthal modes ($K$) for vortices  with $\Omega = 0.25$ and charges 
$m=1,...,5$ (left to right) using the numerical routine
described in Sec.~\ref{SEC:num} to converge the VA ansatz 
into a numerically `exact' solution.  The predictions are made numerically 
integrating the constants of Eq.~(\ref{constants}).  We see that for each 
$m$, after the critical mode, the growth rate predictions for each $K$ 
become $0$ indicating that the perturbations after the critical mode 
are stable.\label{GMprediction}}
\end{figure}

\subsection{Two-dimensional Simulations}
\label{sec_sim}

We now compare our predictions for the MI growth rates for vortex charges
$m=1,...,5$ using Eq.~(\ref{eig2}) to numerical results,
see Fig.~\ref{GMprediction}.
To verify our predictions we use a polar-grid finite-difference scheme where we treat the time derivative separately from the spatial derivatives.
For the time derivatives, we use the fourth order Runge-Kutta method.  For the spatial derivatives we use a second-order central difference scheme:
\[
\nabla^2\Psi_{i,j}= D(\Psi_i)+ \frac{1}{r_i^2}\frac{\Psi_{i,j+1}-2\Psi_{i,j}+\Psi_{i,j-1}}{\Delta\theta^2}.
\]
For our simulations we use $\Omega=0.25$, $\Delta r = 0.35$,
$\Delta \theta = {2\pi}/{80}$, $\Delta t = 0.0005$, with
a length of the simulation $t_{\mbox{\scriptsize max}}=15$,
and a perturbation amplitude $\epsilon = 0.00001$.

\begin{figure}[htb]
\centering
\includegraphics[width=8.5cm]{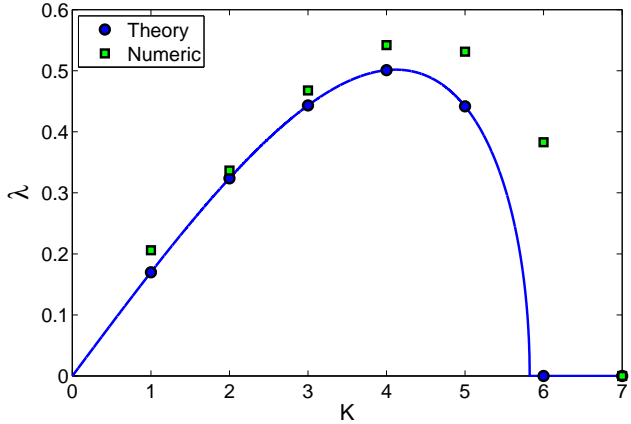}
\vspace{-0.2cm}
\caption[Growth rates from simulation and predictions for $m=2$.]
{(Color online) Average growth rates from full two-dimensional simulation of vortices of charge $m=2$ perturbed with modes $K=1,...,7$ compared to numerical predictions.  The predicted growth rates are shown in blue circles, while the green squares represent the computed growth rates from the full simulation.\label{m2final}}
\end{figure}

\begin{figure}[htb]
\centering
\includegraphics[width=8.5cm]{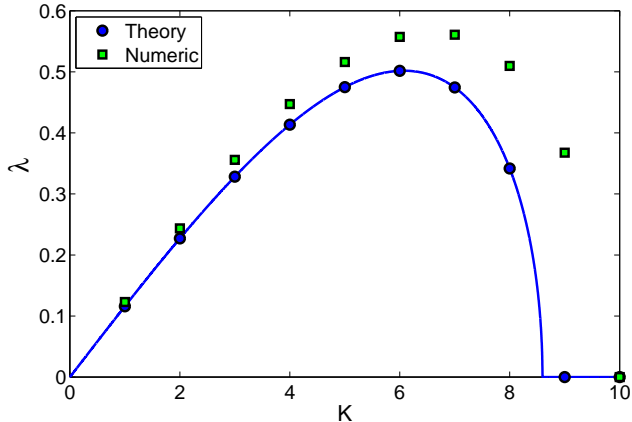}
\vspace{-0.2cm}
\caption[Growth rates from simulation and predictions for $m=3$.]
{(Color online) Same as in Fig.~\ref{m2final} for $m=3$.\label{m3final}}
\end{figure}

Using this scheme, along with Dirichlet boundary conditions,
yields the results in Figs.~\ref{m2final} and \ref{m3final} for
$m=2$ and $m=3$, respectively. The growth rates are calculated by
recording the maximum and minimum of the modulus squared of the
crest of the vortex, and computing the average growth rate of the
perturbation growth.

Overall, we see that our numerical simulations yield growth rates
that are close to those predicted, but typically slightly higher,
with an error on the order of $10\%$ for modes far from
$K_{\mbox{\scriptsize crit}}$.  For modes close to
$K_{\mbox{\scriptsize crit}}$, we observe higher error.  Also, for
$m=2$ and $m=3$, the predicted $K_{\mbox{\scriptsize crit}}$ is
one mode off.

Through one-dimensional simulations, as well as numerical error
analysis, we have accounted for much of this error.  It is
observed that for modes closer to $K_{\mbox{\scriptsize crit}}$,
the simulations are very sensitive to resolution.  By increasing
the resolution to very high levels, the discrepancy in the
one-dimensional runs were virtually eliminated.  Such high
resolutions were not used for the 2D simulations
because the simulations become very computationally expensive.
Additionally, due to the singularities in the $C$-constant
integrals, the numerical predictions derived from them also induce
slight errors.

\begin{figure}[htb]
\includegraphics[width=8.5cm]{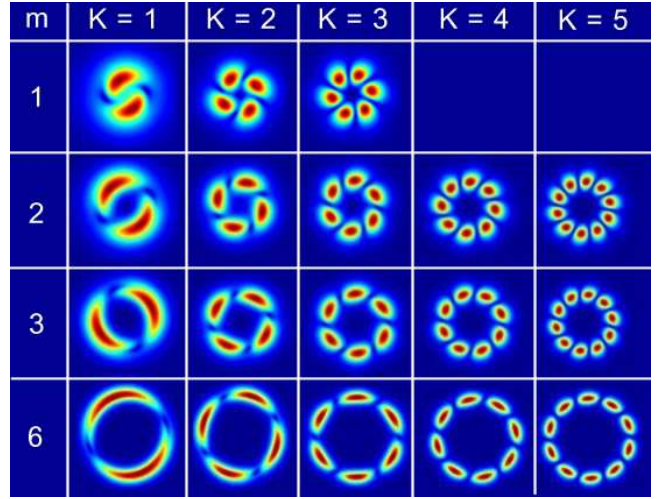}
\caption[Eigenmodes of several vortices.]
{(Color online) Depiction of the modulus squared of numerically derived
unstable eigenmodes of vortices in the 2D focusing NLS of
charges $m=1,2,3,$ and $6$ for modes $K=1,...,5$ (the vortex of charge
$m=1$ does not have unstable modes past $K=3$).  It is obvious from the
panels that the eigenmodes are not completely separable into radial and
azimuthal parts as assumed in Eq.~(\ref{sepAf}), but can be reasonably
approximated by such a separable solution.  We also see that for higher
charges and higher mode numbers, the eigenmodes appear to become more
separable, and thus the approximation of a separable solution becomes
more accurate.\label{nonsepmodesnew}}
\end{figure}

Another source of the discrepancy between our predictions and the
simulations (especially the fact that our critical mode prediction
is off by one) is that the assumption of separability used in
Eq.~(\ref{sepAf}) is not exact, but rather a good approximation.
This can be seen by plotting the 2D eigenvectors of
the steady-state vortices as seen in Fig.~\ref{nonsepmodesnew}.
We see that for low vortex charges, and small mode perturbations,
the eigenvectors are clearly non-separable into radial and
azimuthal parts. As one increases the charge and/or the mode being
perturbed, the eigenvector becomes more separable.  Since our
simulations were done on vortices of low charge, some discrepancy
due to the assumption of Eq.~(\ref{sepAf}) is to be expected
\cite{ronthesis}.

Finally, we note in passing that our approach is somewhat complementary
to the theoretical approach of Ref.~\cite{carr06}, while the combination of
both is in some sense tantamount to the theoretical analog of what
is found numerically in Refs.~\cite{carr06,Herring08}. In Ref.~\cite{carr06}, the
so-called Vakhitov-Kolokolov criterion was considered which is implicitly
connected to the $K=0$ perturbation mode and the instability along
that eigendirection leads to collapse. On the other hand, here we examine
the modulational-type instability of higher $K$ modes which
initiates the unstable dynamics by breakup of the azimuthal symmetry (and
may, however, eventually lead to collapse in conjunction with the
$K=0$ mode, as shown in Fig. \ref{MI_ex}).

\section{Nonlocal Nonlinearity}
\label{sec_nl}

Here we briefly describe one of the possible extensions to the
theory, that of incorporating nonlocal interactions.  Such
interactions correspond to various physical systems, such as
dipole-dipole interactions in a BEC of degenerate dipolar atoms
\cite{NLSnonlocal}, and nonlinear crystals whose nonlinear
refractive index changes due to the intensity of the light present
(determined by a transport process such as heat conduction)
\cite{NLKerr}.
As we will show, the nonlocality of the nonlinearity will induce a stabilizing
effect on the modulational stability of vortices. Other interesting
effects of the nonlocality of the nonlinearity include:
changing, under appropriate circumstances, the character of
interaction of dark solitons from repulsive to
attractive \cite{Kro:06}; changing the interaction
strength between solitons \cite{Kartasho08a};
and stabilization of dipole solitons \cite{Kartasho08b} or
2D ring vortices
such as the ones considered herein \cite{briedis}.
We note that prior work has demonstrated that for the
case of nonlocal $\chi^{(3)}$ nonlinearity, all three dimensional
spatiotemporal solitons with vorticity are unstable \cite{Mihalache}.

%\subsection{Nonlocal Model}
For nonlocal interactions, the NLS can be altered to have a nonlocal
nonlinearity \cite{NLKerr}
\begin{equation}
\label{NLNLS}
i\Psi_t+\nabla^2\Psi+
sN\!\left(\left|\Psi\right|^2\right)\!\Psi=0,
\end{equation}
where the nonlocal nonlinearity takes the form of a convolution
integral:
\[
N
%\left(\left|\Psi\right|^2\right)
=
\int_0^{2\pi}\!\!\int_0^{\infty}
\!V\!\left(r^\prime-r,\theta^\prime-\theta\right)
\left|\Psi\!\left(r^\prime,\theta^\prime,t\right)\right|^2
 r^\prime\,dr^\prime\,d\theta^\prime,
\]
and where $V$, the nonlocal response function, is taken to be
a Gaussian (which appears in relation to the nonlinear crystal
heat diffusion nonlocality \cite{NLKerr}):
\begin{eqnarray}
V\!\left(r'-r,\theta'-\theta \right)
&=&\frac{1}{\pi\sigma^2}
\exp\!\left(-\frac{|\overrightarrow{r}'-\overrightarrow{r}|^2}{\sigma^2}\right),
\notag
\end{eqnarray}
where $\overrightarrow{r}\!=\!(r\cos\theta,r\sin\theta)$ and
$\overrightarrow{r}'\!=\!(r'\cos\theta',r'\sin\theta')$.
%
%\[
%V\!\left(r,\theta \right)
%=\frac{1}{\pi\sigma^2}
%\:\mbox{exp}\!\left[(-r^2\cos^2\theta-
%r^2\sin^2\theta)/\sigma^2\right].
%\]
%
Formulating the Lagrangian density of Eq.~(\ref{NLNLS}) yields:
\[
\mathcal{L}=\frac{i}{2}\left(\Psi\Psi_t^* -
             \Psi^*\Psi_t\right) +
             \left|\Psi_r + \frac{1}{r}\Psi_\theta\right|^2
             - \frac{s}{2}|\Psi|^2N(|\Psi|^2)\left.\right. ,
\]
which, by the same method as in Sec.~\ref{model},
yields the following azimuthal equation of motion:
\begin{equation}
\label{NLAEQ}
i\,C_1A_t = C_2A - C_3A_{\theta\theta} - s\,C(\theta,t)A,
\end{equation}
where $C$ is defined as:
\begin{eqnarray}
\label{Ceq}
C(\theta,t)&=&
\int_0^{2\pi}\!\!\!\int_0^\infty\!\!\!\int_0^\infty
\!V\!\left(r^\prime-r,\theta^\prime-\theta\right)
\,\times \\[1.0ex]
&&
\notag
f(r)^2f(r^\prime)^2|A(\theta^\prime,t)|^2
r r^\prime\,dr\,dr^\prime\,d\theta^\prime.
\end{eqnarray}
Note that the new, nonlocal, azimuthal NLS (\ref{NLAEQ})
is the same as in the local case [see Eq.~(\ref{Aeq0})]
where the (local) $C_4$ integral has been replaced by
the (nonlocal) convolution $C$ integral (\ref{Ceq}).

%\subsection{Nonlocal Stability Analysis}
%\label{nlstable}
We use the same stability analysis technique as in Sec.~\ref{model},
with a slight alteration.  We notice that if we define:
\begin{eqnarray}
R(\theta^\prime-\theta)&=&\int_0^\infty\!\!\int_0^\infty
\!V\!\left(r^\prime-r,\theta^\prime-\theta\right)
\,
\notag
f(r)^2f(r^\prime)^2\,r r^\prime\,dr\,dr^\prime,
\end{eqnarray}
then $C$ is now a convolution term as follows:
\[
C(\theta,t)=
\int_0^{2\pi}\!R(\theta^\prime-\theta)|A(\theta^\prime,t)|^2\,d\theta^\prime=
R*|A|^2.
\]
Inserting this into Eq.~(\ref{NLAEQ}), and using the same rescalings
as in Eqs.~(\ref{Arescale}) and (\ref{trescale}), yields
\[
iA_t=-A_{\theta\theta}-\frac{s}{C_3}A(R*|A|^2).
\]
Now, we perform a stability analysis identical to that of Sec.~\ref{stable}, but along with the transforms of $\hat u$ and $\hat v$, we also add:
\[
\hat R(K)=\int_0^{2\pi}R(\theta)e^{iK\theta},
\]
in which case the convolution term becomes a product ($A\hat R$), and we get:
\[
\lambda_{\pm}=\pm\frac{C_3}{C_1}\sqrt{K^2\left(2s\frac{\hat R(K)}{C_3}-K^2\right)},
\]
and, therefore, the critical mode is:
\[
K_{\mbox{\scriptsize crit}} = \pm \sqrt{2s\frac{\hat R(K)}{C_3}},
\]
which has the same form as the local nonlinearity case after replacing $C_4$ with $\hat R(K)$.  We see that depending on the nonlocal response function, $K_{\mbox{\scriptsize crit}}$ can be damped, and if $\hat R(K)<\left|\frac{C_3}{2s}\right|$ then $K_{\mbox{\scriptsize crit}}<1$ and all modes become stable.  Therefore, we see that one could have a vortex in the focusing NLS with a nonlocal nonlinearity which would be modulationally
{\em stable}. In fact, this modulational stability (as well as the
stability against collapse) of the focusing ring vortices
of the nonlocal NLS equation has been confirmed in the work of
Ref.~\cite{briedis} and is a feature that
could have practical applications, such as data storage and
communications using light vortices in Kerr
optical media \cite{OpticalVortex}.

\medskip
\section{Conclusions\label{SEC:CONCLU}}

We have formulated a methodology for studying {\em azimuthal} modulational instability
of vortices in the two-dimensional 
nonlinear Schr\"odinger (NLS) equation which can be extended to incorporate any additional terms in the NLS as long as they have a Lagrangian representation.  (This expandability of the method adds greatly to its usefulness and
broad relevance). The method relies on approximating a vortex solution as being separable into its radial and azimuthal parts, and using the Lagrangian functional of the NLS to obtain a quasi-one-dimensional equation of motion for the azimuthal direction.  A stability analysis on modulational perturbations of the
equation, leads to predictions of growth rates for each perturbed mode, and of
the critical mode.  After obtaining a steady-state vortex solution using a
variational ansatz along with a nonlinear optimization routine, we ran
numerical simulations of the NLS, perturbing individual modes and
recording their growth rates to confirm the predictions.

One key result that should not be overlooked is that of the usefulness of the
variational ansatz of the vortex profiles that we derived.  Since this profile
seems to converge to the numerically exact solution as the vortex charge
becomes large, experimenters can use it to make simple yet
accurate predictions of the  vortex radius and intensity
given experimental parameters. Furthermore, it can be used as in
Ref.~\cite{carr06}, in both local and nonlocal settings to yield an
approximate threshold for collapse dynamics.

We have also shown theoretical predictions of modulational
instability of vortices which exhibit a nonlocal response by
extending the NLS to incorporate a nonlocal nonlinearity.
The results illustrate that nonlocality can damp, or completely eliminate,
the modulational instability, potentially leading to the complete
stabilization of the nonlocal vortices, as shown numerically, e.g.,
in Ref.~\cite{briedis}.

%\begin{acknowledgments}
\section*{Acknowledgments}
We appreciate valuable discussions with Michael Bromley.
R.M.C., R.C.G., and P.G.K.~acknowledge support from
NSF-DMS-0505663 and NSF-DMS-0806762. P.G.K. also acknowledges
support from NSF-DMS-0349023 and the Alexander von Humboldt Foundation.
%\end{acknowledgments}

\end{document}